\documentclass[prd,showpacs,showkeys,floatfix,nofootinbib, %%,eqsecnum,
               preprint,12pt,tightenlines,fleqn]{revtex4} %for preprint
\usepackage{amssymb,epsf}
\usepackage{latexsym}
\usepackage{textcase,bm}
\usepackage{amsmath,mathrsfs}
\usepackage{graphicx,epsfig}
\usepackage{latexsym,amssymb}
\usepackage[utf8x]{inputenc}
\usepackage{graphics}
\usepackage[english]{babel}
\usepackage{graphicx}
\usepackage{subfig}

\newcommand  {\version}{v3}  %%true version v2.992  %%September 6, 2012

\newcommand{\beq}{\begin{equation}}
\newcommand{\eeq}{\end{equation}}
\newcommand{\beqa}{\begin{eqnarray}}
\newcommand{\eeqa}{\end{eqnarray}}
\newcommand{\bsubeqs}{\begin{subequations}}
\newcommand{\esubeqs}{\end{subequations}}
\begin{document}
%
%\preprint{KA--TP--34--2012}\;(\today;\;\version)
%
\noindent  arXiv:1208.3168
\hfill KA--TP--34--2012\;(\version)\newline\vspace*{2mm}
\title{\vspace*{2mm}Gravity Without Curved Spacetime:
A Simple Calculation\\ \vspace*{2mm}}
\author{F.R.~Klinkhamer}
\email{frans.klinkhamer@kit.edu}\affiliation{Institute for
Theoretical Physics, Karlsruhe Institute of
Technology (KIT), 76128 Karlsruhe, Germany\\}

\begin{abstract}%
\noindent %\vspace*{0mm}\newline
Classical-particle trajectories are calculated for the
static Einstein universe without requiring
that the 3-space be closed and curved.
Freely-moving test particles are found to return to their
starting positions because of strong gravitational-field effects.
Possible implications for the underlying (quantum) theory
of gravity are briefly discussed.
\end{abstract}



\maketitle

\section{Introduction}
\label{sec:Introduction}

In the decades leading up to the construction of
the standard model of elementary particle physics,
the idea was explored that gravity could also be viewed as a
(spin--2) gauge field theory in Minkowski spacetime.
A selection of original articles/lectures is given in
Refs.~\cite{Kraichnan1955,Gupta1957,Thirring1961,Feynman1963,%
Weinberg1965,Deser1970,Veltman1976}
and a selection of related textbook discussions
in Refs.~\cite{Feynman1964,{Weinberg1972},MTW1973}.

The conclusion reached was that there is no fundamental
difference between this  flat-spacetime gauge-field-theory point of view
and the standard geometric point of view (the one adopted by
Einstein~\cite{Einstein1916,Einstein1917}).
See, e.g., Sec.~8.3 of Ref.~\cite{Feynman1963} and
Box 18-1 of Ref.~\cite{MTW1973}.

However, the naive question arises as to how the flat-spacetime approach
would be able to describe, for example, a closed cosmological model.
Veltman and the present author addressed this puzzle
some time ago~\cite{KlinkhamerVeltman1992}
but did not reach a definite conclusion,
because two contradictory results were obtained for the
particle trajectories. Now, this contradiction
has been resolved with the realization
that the first of these results does not apply globally,
whereas the second does.

We describe the outcome of this
simple (and even trivial) calculation
in Secs.~\ref{sec:Gravity}--\ref{sec:Orbits-in-flat-space}.
At the end of this article, in Sec.~\ref{sec:Conclusion},
we point out that the result may be especially relevant
if gravity is not fundamental but emergent.

%%\newpage%%tmp
\section{Gravity}
\label{sec:Gravity}

As explained in the Introduction, the starting point is
the suggestion~\cite{Kraichnan1955,Gupta1957,Thirring1961,%
Feynman1963,Weinberg1965,Deser1970,Veltman1976}
that gravity is a non-Abelian gauge field theory
of a massless spin--2 field $h_{\mu\nu}(x)$ defined over
flat Minkowski spacetime, just as the electromagnetic and strong
interactions of the standard model are
non-Abelian gauge field theories of spin--1 fields over
Minkowski spacetime. In short, this description
of gravity does not rely on the curved-spacetime concept.

Purely for convenience, define
\bsubeqs
\begin{equation}\label{eq:metric}
g_{\mu\nu}(x)\equiv \eta_{\mu\nu}+h_{\mu\nu}(x)\,,
\end{equation}
where $\eta_{\mu\nu}$ is the Minkowski metric, here taken as
\begin{equation}\label{eq:minlowski-metric}
\eta_{\mu\nu}=\text{diag}(-1,\,-1,\,-1,\,+1)\,.
\end{equation}
\esubeqs
A finite gauge transformation with parameters $\xi^{\kappa}(x)$
then transforms the spin--2 field $h_{\mu\nu}(x)$
contained in $g_{\mu\nu}(x)$
and a generic scalar field $\phi(x)$ as follows:
\bsubeqs
\beqa
 g_{\mu\nu}\big(x\big) &\rightarrow&
 \frac{\partial\xi^{\kappa}}{\partial x^{\mu}}
 \frac{\partial\xi^{\lambda}}{\partial x^{\nu}}\;
g_{\kappa\lambda}\big(\xi (x)\big)\,,
\\[2mm]
\phi\big(x\big)  &\rightarrow&  \phi\big(\xi (x)\big)\,.
\eeqa
\esubeqs
For the gravitational coupling of matter, gauge invariance
effectively plays the role of the equivalence principle
(cf. the last paragraph of Sec.~13 in Ref.~\cite{Veltman1976}).
In fact, the invariant action (units $8\pi G_{N} = c = 1$) reads
\begin{equation}\label{eq:action}
A=\int_{\mathbb{R}^4} d^4 x\;
\big[-\textstyle{\frac{1}{2}}\,\sqrt{g(x)}\,R(x)
+\sqrt{g(x)}\,\mathcal{L}_m(x)\big],
\end{equation}
with the density $g(x)$ given by the determinant of the
$4 \times 4$ matrix \eqref{eq:metric}
and the kinetic term $R$ equal to the
Ricci ``curvature'' scalar in terms of the
``metric'' \eqref{eq:metric}.
Note that the spacetime integral
in \eqref{eq:action} is always over $\mathbb{R}^4$.

The standard geometric point of view
(developed by Einstein~\cite{Einstein1916})
is that there exist rigid measuring-rods and perfect standard-clocks
in a curved space-time with line-element
$ds^2=g_{\mu\nu}\,dx^{\mu}dx^{\nu}$.
The alternative point of view (discussed by
Feynman~\cite{Feynman1963}, for example) is that the
rods and clocks are deformed by a gravitational field $h_{\mu\nu}(x)$
defined over Minkowski spacetime.
Feynman~\cite{Feynman1963}
and Misner--Thorne--Wheeler~\cite{MTW1973}, amongst others,
conclude that there are no important
differences between these two points of views.

Expanding on these somewhat abstract discussions,
we adopt a pragmatic approach and consider
the trajectories of a classical test particle, in particular, the
paths in a ``closed'' universe.

%%\newpage%%tmp
\section{Equation of motion}
\label{sec:Equation-of-motion}

As pointed out by Einstein and Grommer (1927),
it is possible to obtain the motion of a classical
particle already from the gravitational field equations;
see, e.g., Sec.~20.6 of Ref.~\cite{MTW1973}
and Sec.~11.1 of Ref.~\cite{AdlerBazinSchiffer1975}
for further discussion and references.
A simple derivation of the equation of motion of a classical test particle
with position $x_{p}$ then starts from energy-momentum conservation,
\begin{equation}
T^{\mu\nu}_{\phantom{\mu\nu};\nu}=0\,.
\end{equation}
Consider dust particles with
\bsubeqs%%\esubeqs
\begin{eqnarray}
T^{\mu\nu} &=& \rho_0(x)\,u^\mu(x)u^\nu(x)\,,
\\[2mm]
u^\mu &=& \frac{dx^\mu}{ds}\,,
\quad
u^\mu u_\mu =1\,.
\end{eqnarray}
\esubeqs
It now follows that~\cite{AdlerBazinSchiffer1975}
\bsubeqs%%\esubeqs
\begin{equation}\label{eq:geodesic-equation}
 \frac{d^2x_{p}^{\mu}}{ds^2}+\Gamma^{\mu}_{\kappa\lambda}(x_{p})\,
 \frac{dx_{p}^{\kappa}}{ds}\frac{dx_{p}^{\lambda}}{ds}=0\,,
\end{equation}
with the standard affine connection
given in terms of the ``metric'' \eqref{eq:metric}, its inverse,
and its derivatives,
\beq\label{eq:Gamma}
\Gamma^{\mu}_{\kappa\lambda} \equiv
\frac{1}{2}\,g^{\mu\nu}\,\left(
 \frac{\partial g_{\nu\kappa}}{\partial x^{\lambda}}
+\frac{\partial g_{\nu\lambda}}{\partial x^{\kappa}}
-\frac{\partial g_{\kappa\lambda}}{\partial x^{\nu}}
\right)\,.
\eeq
\esubeqs
The equation of motion  \eqref{eq:geodesic-equation} is better known as
the ``geodesic equation,'' here with quotations marks because we
do not adopt a geometric point of view (or, at least, do not start from it).

Equation \eqref{eq:geodesic-equation}
is form-invariant under gauge transformations,
\bsubeqs\label{eq:gauge-transformation}
\begin{eqnarray}
g_{\mu\nu}(x) &\rightarrow& \frac{\partial\xi^{\kappa}}{\partial x^{\mu}}
\frac{\partial\xi^{\lambda}}{\partial x^{\nu}}\;g_{\kappa\lambda}(\xi )\,,
\\[2mm]
x^\mu_{p} &\rightarrow& \xi^\mu(x_{p})\,,
\end{eqnarray}
\esubeqs
as long as the gauge parameter $\xi^\kappa = \xi^\kappa(x)$
is an invertible function.

%%\newpage%%tmp
\section{RW Universe: $\boldsymbol{k=1}$ case}
\label{sec:RW Universe}

For definiteness, consider a specific gravitational background field
corresponding to the Robertson-Walker (RW)
metric~\cite{AdlerBazinSchiffer1975,RobertsonNoonan1968}
with constant positive ``curvature'' of the spatial hypersurfaces
($k=1$ in the standard terminology).
With Cartesian coordinates $x^{i}$ of the 3-space and
the cosmic scale factor $a=a(t)$, the gravitational background field
\eqref{eq:metric} is given by
\bsubeqs\label{eq:RW}
\beqa
g_{\mu\nu}(x^{1},\,x^{2},\,x^{3},\,t) &=&
\begin{pmatrix}
 &                           &   &\;\;\;0\\
 &  -a(t)^2\,\widetilde{g}_{ij}(x) &   &\;\;\;0\\
 &                           &   &\;\;\;0\\
0 & 0                        &  0&\;\;\;1
\end{pmatrix}\,,
\\[2mm]
\label{eq:RW-gtilde}
\widetilde{g}_{ij}(x^{1},\,x^{2},\,x^{3}) &=&
\left(\frac{1}{1+k\,|\vec{x}|^2/4}\right)^{2}\;\delta_{ij}\,,
\\[2mm]
\label{eq:RW-k=1}
k&=&1 \,.
\eeqa
\esubeqs
There are no singularities in \eqref{eq:RW} and
one coordinate patch suffices for $|\vec{x}| < \infty$.
Note that the spatial coordinates $x^{i}$ are dimensionless
and the cosmic scale factor $a(t)$ has the dimension of length,
just as the cosmic time coordinate $t$ (recall $c=1$).

It is, then, easy to evaluate the equation of motion
\eqref{eq:geodesic-equation} using
%\bsubeqs\label{eq:RW-Gammas}
%\begin{eqnarray}
\beq\label{eq:RW-Gammas}
%%%FRK these Gamma's same as W(15.1.3-5) --> standard FRW eqs.
\Gamma^i_{oj} = \frac{\dot{a}}{a}\;\delta^i_j\,,
\quad %%\\[2mm]
\Gamma^0_{ij} = \dot{a}\,a\;\widetilde{g}_{ij}\,,
\quad %%\\[2mm]
\Gamma^i_{jk} =\widetilde{\Gamma}^i_{jk}\,,
\eeq
%\end{eqnarray}
%\esubeqs
where the overdot stands for differentiation with respect
to $t$ and $\widetilde{\Gamma}^i_{jk}$ is the 3-dimensional version of
\eqref{eq:Gamma} for the metric \eqref{eq:RW-gtilde}.
The results for the 3-dimensional trajectories
($du^2=\widetilde{g}_{ij}\,dx^{i}dx^j$) are as follows:
\bsubeqs\label{eq:RWclosed-geodesic}
\begin{equation}\label{eq:RWclosed-geodesic-3d}
\frac{d^2x^{i}}{du^2}
+\widetilde{\Gamma}^{i}_{jk}\,\frac{dx^{j}}{du}\frac{dx^{k}}{du}=0\,,
\end{equation}
\begin{equation}\label{eq:RWclosed-geodesic-parameter}
\left(\frac{dt}{du}\right)^2 =a^2+\text{const}\,.
\end{equation}
\esubeqs

It suffices for our purpose to consider the static Einstein
universe~\cite{Einstein1917} with a
pressureless perfect fluid as matter component
and a cosmological constant $\Lambda>0$
(units $8\pi G_{N} =c= 1$):%
\bsubeqs\label{eq:static-Einstein-universe}
\begin{eqnarray}
\label{eq:static-Einstein-universe-a}
a &=& 1/\sqrt{\Lambda}\,,
\\[2mm]
\label{eq:static-Einstein-universe-rhoM}
\rho_{m} &=& 2\,\Lambda\,,
\\[2mm]
\label{eq:static-Einstein-universe-PM}
P_{m} &=& 0\,,
\end{eqnarray}
\esubeqs
which solves the reduced Einstein field equations, i.e., the
standard Friedmann equations obtained by inserting
the expressions \eqref{eq:RW-Gammas}.
[A similar discussion holds for the global de Sitter metric
(cf. Sec.~16.3 of Ref.~\cite{RobertsonNoonan1968}),
which is also given by
\eqref{eq:RW} but now with the time-dependent scale factor
$a(t) = \sqrt{3/\Lambda}\, \cosh(\sqrt{\Lambda/3}\; t)$.]

From \eqref{eq:RWclosed-geodesic-parameter}
and \eqref{eq:static-Einstein-universe-a}, we have
$u\propto t$, depending on the initial velocity of the particle,
and we only need to solve for the 3-dimensional orbits from
\eqref{eq:RWclosed-geodesic-3d}.

%%\newpage%%tmp
\section{Orbits in flat space}
\label{sec:Orbits-in-flat-space}

\begin{figure*}[b]%%[b]%%%%[p]
\begin{center}                       %%fig1tmp4=fig1tmp5=fig1_v1=fig1_v3
\includegraphics[width=7cm]{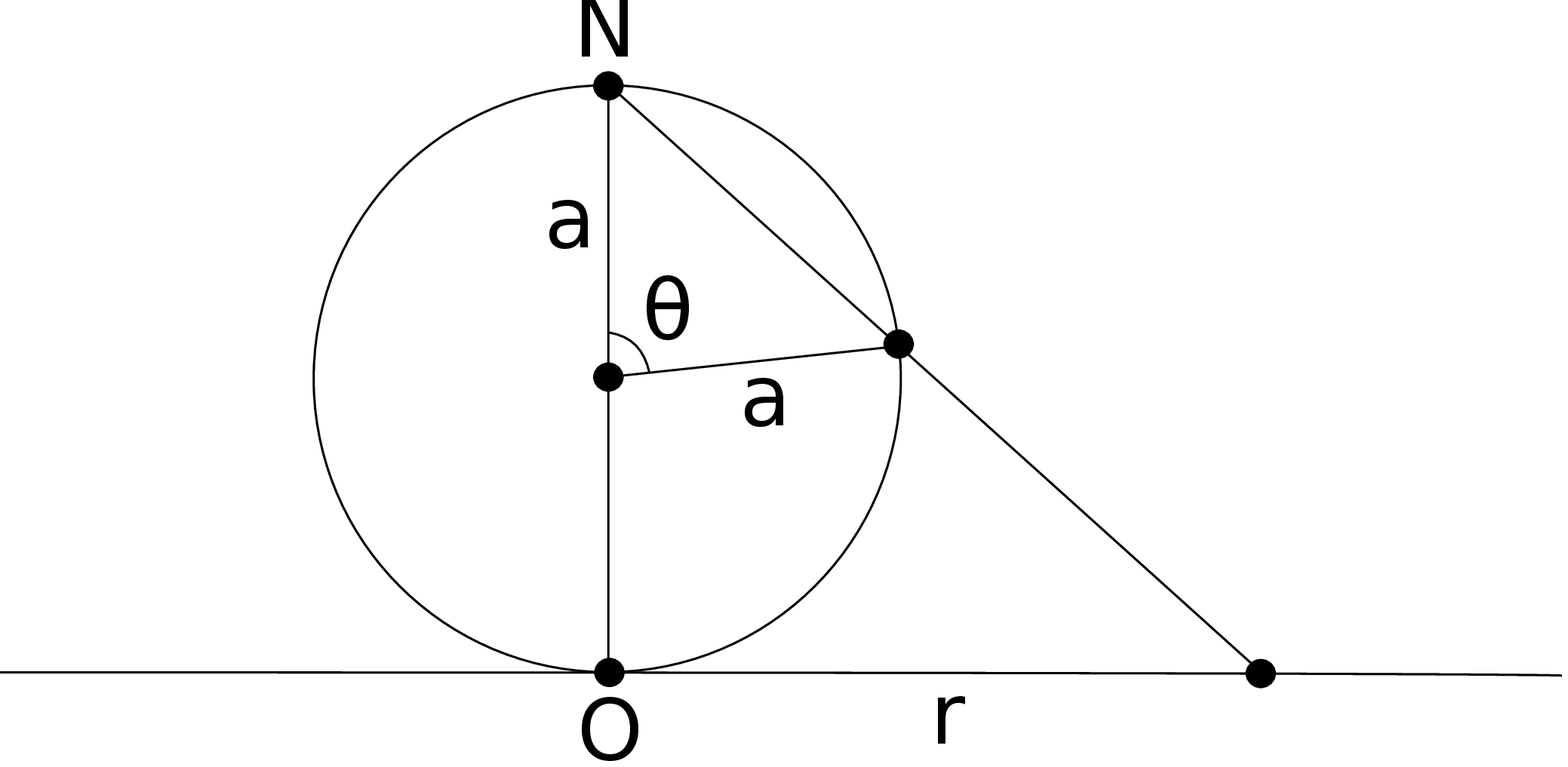}
\end{center}
\vspace*{-0.5cm}
\caption{\noindent
Coordinates for the physical plane and the auxiliary sphere, see main text.}
\label{fig:one}
\begin{center}                               %%fig2tmp5=fig2_v1==fig2_v3
\includegraphics[width=7cm]{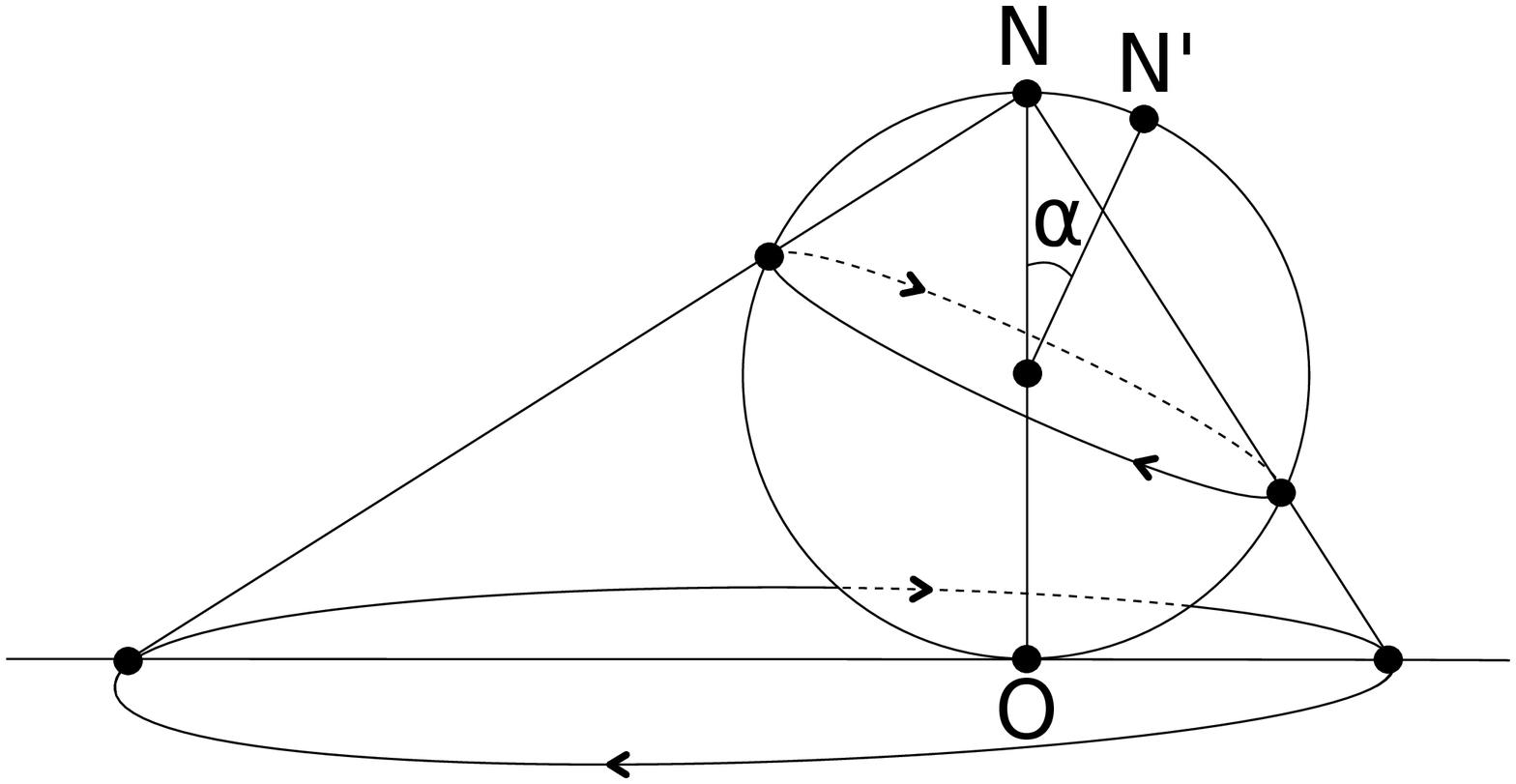}
\end{center}
\vspace*{-0.5cm}
\caption{\noindent Mathematical construction of the particle trajectories
in the plane, see main text.}
\label{fig:two}
\end{figure*}

Consider, first, the 2-dimensional case by
taking the $x^{3}=0$ slice of the space manifold $\mathbb{R}^3$.
Next, change the remaining two coordinates as follows:
\bsubeqs
\begin{eqnarray}
\theta &=&
\pi-2\,\arctan\left(\frac{\sqrt{(x^{1})^2+(x^{2})^2}}{2}\right)\,,
\\[2mm]
\phi  &=& \arctan\left(\frac{x^{2}}{x^{1}}\right)\,,
\end{eqnarray}
which gives the following expression for the
2-dimensional metric:
\begin{eqnarray}
a^2\,\widetilde{g}_{ij}\,dx^{i}dx^j &=&
a^2\,\big(d\theta^2+\sin^2\theta\, d\phi^2\big)\,,
\label{eq:RW-gtilde-sphere}
\end{eqnarray}
\esubeqs
where the indices $i,j$ are summed over $1$ and $2$.
From the expression  \eqref{eq:RW-gtilde-sphere} we
are led to the following
\emph{mathematical trick}:
the 2-dimensional gravitational field from \eqref{eq:RW-gtilde}
can be \emph{interpreted} as the metric on a 2-sphere with radius $a$.
See Fig.~\ref{fig:one}, where $r$ is the radial coordinate from the
point O in the plane and $\phi$ (not shown) is the azimuthal angle.

It is now easy to obtain the orbits in the
flat 2-dimensional space~\cite{RobertsonNoonan1968}: they are
simply the projections of great circles on the sphere
(each with a pole $\text{N}'$
having spherical coordinates $\theta =\alpha$, $\phi =\beta$;
see Fig.~\ref{fig:two}).
The resulting orbits in the plane turn out to be circles with
the following center positions and radii:
\bsubeqs\label{eq:trajectories}
\beqa
\big(r,\,\phi\big)_\text{center}
&=&
\big(2\,a\,\tan\alpha,\,\beta \big)\,,
\\[2mm]
l_\text{radius}&=&\frac{2\,a}{\cos\alpha}\,.
\eeqa
\esubeqs

\begin{figure*}[t]%%[t]%%[p]
\begin{center}%%\end{center}  %%
\includegraphics[width=7cm]{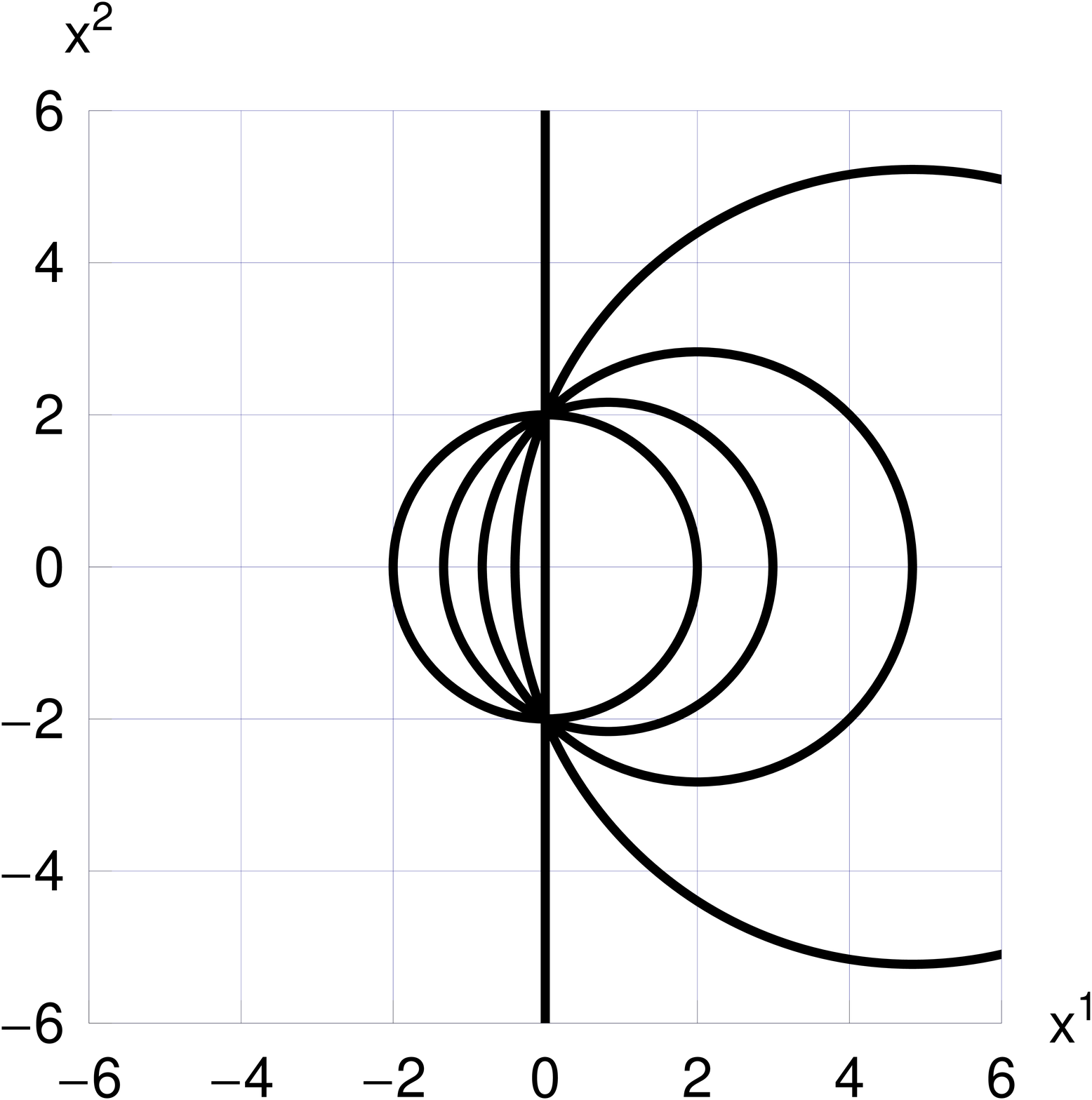}
\end{center}
\vspace*{-0.5cm}
\caption{\noindent Selected particle trajectories in
the static Einstein universe \eqref{eq:static-Einstein-universe}
with $\Lambda=1$. These trajectories
can be obtained mathematically by projecting great
circles with polar angle
$\alpha$ $=$ $(0,\,1,\,2,\,3,\,4-0^{+})$ $\times$ $\pi/8$
as defined in Fig.~\ref{fig:two} or
by directly solving the equation of motion
\eqref{eq:RWclosed-geodesic-3d} with appropriate boundary conditions.
Further trajectories are obtained by reflection
$x^{1}\to -x^{1}$.%\vspace*{15cm}
}
\label{fig:three}
\end{figure*}

A ``straight'' path along the $x^{2}$ axis, for example, is obtained
by considering a particular great circle in Fig.~\ref{fig:two},
which passes through O (i.e., $\alpha=\pi/2$)
and is traversed many times in the
same direction. The corresponding particle motion in the plane
(i.e., the real space according to our point of view)
then always passes O from the \emph{same} direction,\footnote{An
earlier result (Part I of Ref.~\cite{KlinkhamerVeltman1992})
suggested that the particle would return to the starting
point from \emph{alternating} directions. But, as mentioned in the
Introduction, we now realize that this is not the case,
because that particular solution of the equation of motion
does not apply at the boundary of the coordinate patch used.}
because the path
``closes at infinity,'' where $\widetilde{g}_{ij}=0$.
Indeed, consider a great circle with
$\alpha=\pi/2-\epsilon$ and let $\epsilon\to 0^{+}$.
In the plane, this gives a path along the $x^{2}$ axis
closed by a particular semi-circle ``at infinity''
(the other semi-circle occurs
for $\epsilon\to 0^{-}$). The behavior is shown in
Fig.~\ref{fig:three}, where the circle segments in the plane get closer
and closer to the $x^{1}=0$ line as $\alpha$ increases towards
the value $\pi/2$.

Let us emphasize that the particle trajectories of
Fig.~\ref{fig:three} are the paths taken by classical test
particles in the real physical flat space (here, the slice
$x^{3}=0$): all previous discussions of great circles
and projections are purely mathematical short-cuts.
The same paths follow directly from the equation of motion
\eqref{eq:RWclosed-geodesic},
starting at, for example, $(x^{1},\,x^{2})=(0,\,2)$
and having initial velocities directed towards the upper right
or the lower left in Fig.~\ref{fig:three}.

%%\newpage%%tmp
Several remarks are in order. First, the
orbits for the 3-dimensional case lie in the plane defined by
the initial position $\vec{x}_{start}$
and the initial velocity $\vec{v}_{start}$.
In that plane, the discussion is the same as for the
2-dimensional case.
We continue, therefore, with the 2-dimensional case.

Second, all orbits \eqref{eq:trajectories}
have the same proper length $2\pi a$,
as follows from the identification \eqref{eq:RW-gtilde-sphere}
and the construction of Fig.~\ref{fig:two}.
The orbits cross because of tidal effects of the gravitational field.

Third, the ``center of the universe'' O in the plane
of Fig.~\ref{fig:two},
around which the particles circle,
is a gauge artifact. A gauge transformation \eqref{eq:gauge-transformation}
with $\xi^\mu(x)=x^\mu+b^\mu$ moves it to an arbitrary position,
\bsubeqs
\begin{eqnarray}
x^\mu =0 &\rightarrow& \xi^\mu(0)=b^\mu\,,
\end{eqnarray}
and corresponds to an isometry,
\begin{eqnarray}
g_{\mu\nu}(x) &\rightarrow& g_{\mu\nu}'(\xi)=g_{\mu\nu}(\xi)\,.
\end{eqnarray}
\esubeqs
The reader is referred to, e.g., Sec.~13.1 of
Ref.~\cite{Weinberg1972} for a general discussion of isometries.

Fourth, any circle in the 3-space with metric
\eqref{eq:RW-gtilde} and \eqref{eq:RW-k=1}
has the following ratio of proper distances:
\begin{equation}\label{eq:circle-radius-circumference}
\frac{d^\text{circumference}}{d^\text{radius}}=
2\pi\;\frac{\sin (d^\text{radius}/a)}{d^\text{radius}/a}\,.
\end{equation}
From the flat-spacetime point of view discussed
in the penultimate paragraph of Sec.~\ref{sec:Gravity}, the behavior
\eqref{eq:circle-radius-circumference} is interpreted
as being due to changing measuring-rods
(cf. the bug on the hot plate in Figs.~42-2 and 42-12
of Ref.~\cite{Feynman1964})
rather than having a curved space with constant measuring-rods
(cf. Sec.~3 of Ref.~\cite{Einstein1916}).

%%\newpage%%tmp
\section{Conclusion}
\label{sec:Conclusion}

The 3-space of the static Einstein universe
($\rho_{m}=2\,\Lambda>0$ and $P_{m} = 0$)
can be considered to be flat ($M_3=\mathbb{R}^3$) and to have
gravitational fields which  become strong far out
 ($h_{ij} \rightarrow \delta_{ij}$  for $|\vec{x}|\rightarrow\infty$),
shrinking the measuring-rods and making the physical distances
vanish at ``infinity.''
This effectively turns $\mathbb{R}^3$ into $S^3$ and there is,
to paraphrase Wheeler~\cite{Wheeler1968}, ``topology without topology.''
A simple calculation thus provides
an example of how flat space can mimic a closed space, by having
particles return in a finite time due to strong gravitational fields.

For classical gravity there is apparently no difference between
the geometric and gauge-field-theory
interpretations, but, most
likely, this does not hold at the quantum level.
Indeed, already for quantum field theory
without gravity ($G_{N}=0$), it matters if $M_3$ equals
$\mathbb{R}^3$ or $S^3$, as exemplified by the so-called
CPT anomaly (a boundary-condition
effect for chiral gauge theories) which is present for the 3-torus $T^3
\subset \mathbb{R}^3$ but not for
the 3-sphere $S^3$~\cite{Klinkhamer2000}. \textit{A
fortiori}, it may turn out to be important
for a future theory of ``quantum gravity'' that the empty spacetime to be
quantized is flat and topologically trivial. The physical (quantum) vacuum
is, of course, known to be far from empty and the cosmological constant
problem still requires a solution.

In this respect, it is to be noted that
the discussion of the present article is directly
relevant to an emergence scenario for the origin of
gravity~\cite{Bjorken2001,Laughlin2003,FroggattNielsen2005,Volovik2008},
which suggests a compensation-type solution of the
cosmological constant problem~\cite{KlinkhamerVolovik2008}.
More specifically, the present article corrects
a previous statement (last paragraph in Ref.~\cite{KlinkhamerVolovik2005}),
which claimed that the existence of a spatially closed universe
would rule out the hypothesis
of gravity emerging from flat spacetime.
It is now clear that spacetime may ``really'' be flat
and still ``appear'' to be closed (cf. Fig.~\ref{fig:three}).
In turn, this implies that an emergence origin of gravity is not
ruled out by a geometric argument.

\section*{\hspace*{-4.5mm}ACKNOWLEDGMENTS}
\vspace*{-0mm}\noindent
It is a pleasure to thank V. Emelyanov and S.~Thambyahpillai for
help with the preparation of this article
and M. Veltman and G.E. Volovik for valuable discussion over the years.

%%\newpage%%tmp
%\vspace*{10cm}

\end{document}